# Adapting Segment Anything Model 3 for Concept-Driven Lesion Segmentation in Medical Images: An Experimental Study

Guoping Xu[1], Jayaram K. Udupa[2], Yubing Tong[2], Xin Long[3], Ying Zhang[1], Jie Deng[1], Weiguo Lu[1], You Zhang[1#]

[1]The Medical Artificial Intelligence and Automation (MAIA) Laboratory, Department of Radiation Oncology, University of Texas Southwestern Medical Center, Dallas, TX 75390, USA

[2]Medical Image Processing Group, Department of Radiology, University of Pennsylvania, Philadelphia, PA 19104, USA

[3]Division of Digestive and Liver Diseases, Department of Internal Medicine, University of Texas Southwestern Medical Center, Dallas, TX 75390, USA

[#] Correspondence: you.zhang@utsouthwestern.edu

**Abstract**

Accurate segmentation of lesions is fundamental to medical image analysis, yet most existing approaches are designed for specific anatomical sites and imaging modalities, limiting their applicability in heterogeneous clinical environments. Recent vision-language foundation models have enabled concept-driven segmentation in natural images, offering a promising path toward more flexible and generalizable medical image segmentation. However, to our knowledge, concept-prompt-based lesion segmentation in medical images, particularly using the latest Segment Anything Model 3 (SAM3), has not been thoroughly examined.

In this work, we present a systematic investigation of the SAM3 framework for lesion segmentation. We first conduct an extensive evaluation of SAM3 using geometric bounding boxes as prompts, as well as concept-based text and image exemplars, for lesion segmentation across multiple cancer types and imaging modalities, including multiparametric MRI, CT, ultrasound, dermoscopy, and endoscopic imaging. To improve robustness and segmentation accuracy, we further explore the integration of additional knowledge into the input representation, including the predictions from adjacent slices, the multiparametric imaging information, and prior annotations. In addition, we provide a comprehensive comparison of fine-tuning strategies for SAM3, encompassing partial module fine-tuning, adapter-based optimization, and full-model fine-tuning under different training objectives.

Extensive experiments conducted on 13 datasets covering 11 lesion types demonstrate that SAM3 achieves strong cross-modality generalization, robust concept-prompt-based segmentation, and accurate instance-specific lesion delineation. These results highlight the potential of concept-driven foundation models for scalable and practical medical image segmentation in real-world clinical settings. Code and trained models will be made publicly available at https://github.com/apple1986/lesion-sam3

# 1. Introduction

Medical image segmentation aims to accurately identify and delineate anatomical structures, tissues, or pathological lesions across diverse imaging modalities. It plays a critical role in numerous clinical applications, including diagnosis, treatment planning, disease monitoring, and quantitative assessment of pathological changes [1, 2]. In recent years, the rapid development of deep learning has led to substantial performance improvements in medical image segmentation, enabling increasingly accurate and automated analysis of complex medical imaging data.

Generally, existing deep learning-based medical image segmentation methods have adopted different strategies to improve the segmentation accuracy and robustness: (1) Multi-scale feature representation: this strategy is widely adopted to handle large variations in object size and is typically achieved through skip connections (e.g., U-Net [3] and UNet++ [4]), enlarged receptive fields via dilated convolutions or large kernels (e.g., the DeepLab series [5, 6] and SegNeXt [7]), or hierarchical feature pyramids constructed using pooling or strided convolutions (e.g., PSPNet [8] and UPerNet [9]); (2) Local and global feature fusion: hybrid architectures that combine convolutional neural networks with Transformer-based modules have been proposed to jointly capture fine-grained local details and long-range contextual dependencies, as exemplified by TransUNet [10], LeViT-UNet [11], and Swin-UNet [12]; (3) Prior-knowledge enhancement: prior-knowledge integration has been explored to further improve segmentation performance, including extending 2D slice-based segmentation to 3D volume-based frameworks to exploit spatial continuity and shape information (e.g., UNETR [13] and UNETR++ [14]), employing advanced data augmentation strategies [15], and introducing deep supervision by injecting label information at multiple stages of the decoder [16]; and (4) other complementary approaches: additional methods investigate novel loss functions [17], uncertainty modeling [18], semi-supervised learning [19], and related techniques to further enhance robustness and generalization in challenging clinical scenarios.

While these approaches or techniques demonstrate impressive segmentation performance, they are often tailored to specific anatomical structures within a fixed body region (e.g., brain, thorax, or abdomen) and a single imaging modality, like MR or CT. Consequently, such models typically require retraining or extensive fine-tuning when applied to different anatomical sites or imaging modalities. As a result, they often struggle to generalize effectively across the diverse and heterogeneous clinical environments encountered in real-world medical practice [20]. Driven by the remarkable success of large language models (LLMs), recent research has highlighted the effectiveness of scaling laws, wherein training Transformer-based architectures with billions of parameters on massive datasets leads to substantial improvements in generalization and task adaptability [21, 22]. Inspired by these advances, large-scale vision foundation models have been actively investigated by leveraging unprecedented amounts of training data, increased model capacity, and novel training strategies to learn more discriminative visual representations [23, 24].

In the context of image segmentation, vision foundation models aim to provide a unified and transferable framework capable of adapting to diverse segmentation tasks with high generalization ability and robustness in real-world scenarios. A representative example is the Segment Anything Model (SAM) introduced in 2023 [25]. Trained on a large-scale dataset comprising 11 million images and over 1 billion masks, SAM demonstrates remarkable generalization capability and strong segmentation performance on natural images under a prompt-based segmentation paradigm. Building upon its architecture, subsequent studies have extended SAM to the medical imaging domain through large-scale medical image pretraining, including MedSAM for 2D medical image segmentation [26] and SAM-Med3D for volumetric data [27]. In 2024, its successor, SAM2, was proposed [28], which extends promptable segmentation from static images to videos. By incorporating streaming memory modules to leverage features and predictions from previous frames, SAM2 enables efficient and real-time multi-object video segmentation using simple geometric prompts, such as points, bounding boxes, or scribble masks. Further studies, including DD-SAM2 [29] and TSMS-SAM2 [30], fine-tuned SAM2 on medical video datasets and demonstrated its strong performance in various clinical scenarios.

Despite these advances, existing SAM-based segmentation approaches remain limited to visual-prompt-driven interaction and cannot segment object instances using a concept-level prompt, like text (short phrase) or an image exemplar. In other words, these interactive models require users to manually provide geometric prompts (e.g., points or bounding boxes) for each target instance on an image, which becomes inefficient or impractical in scenarios involving multiple instances. When multiple instances of the same kind of object appear in a single image, geometric prompts must be provided individually for each instance. To address this limitation, SAM3 [31] was proposed, which not only retains compatibility with visual prompts as in SAM and SAM2, but enables concept-based segmentation, allowing all instances corresponding to a given concept prompt, which is specified by text or an image exemplar, to be segmented simultaneously (see **Fig. 1** for an illustrative comparison between geometric-based and concept-based prompting in SAM3. Note that geometric-based prompts yield a single tumor segmentation per prompt, whereas concept-based prompting enables simultaneous segmentation of all relevant tumor regions).

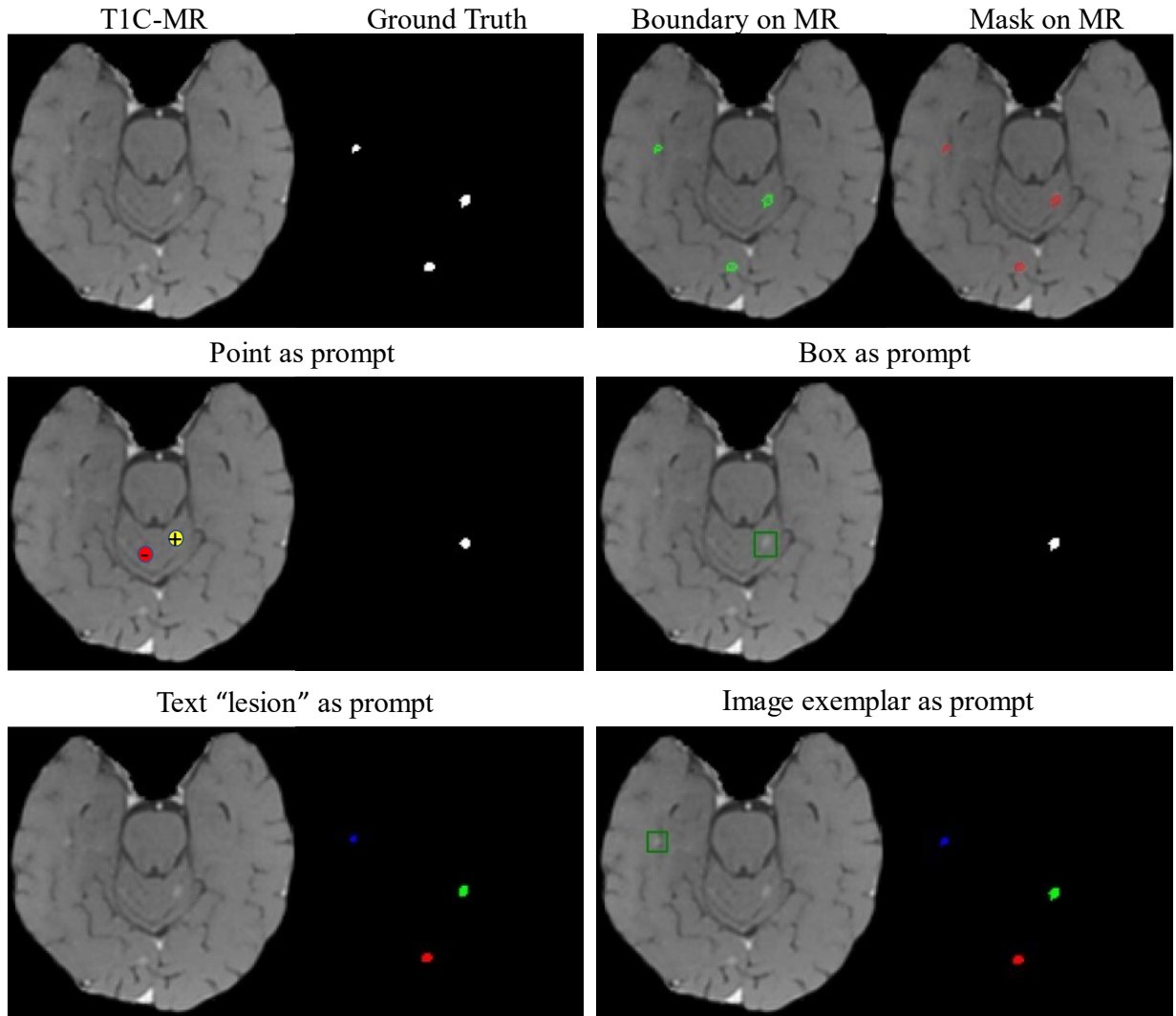

**Figure 1.** Comparison of geometric-based and concept-based prompts in SAM3. The first row shows a T1-weighted contrast-enhanced MRI slice, its tumor annotation (3 lesions), tumor boundary overlay, and tumor mask overlay. The second row presents geometry-based prompting using point prompts (positive in yellow, negative in red) or a bounding-box prompt, together with the corresponding segmentation results. The third row illustrates concept-based segmentation using either a text prompt ("lesion") or an image exemplar (green bounding box) and the resulting prediction masks. Geometry-based prompts yield a single tumor prediction per prompt, whereas concept-based prompting enables simultaneous segmentation of all relevant lesion regions.

Recent studies have begun to explore the potential and limitations of SAM3 by analyzing its architectural and functional advantages compared with SAM2. Specifically, the work in [32] investigates the fundamental differences between SAM2 and SAM3 on natural images, aiming to elucidate why SAM2 fails in concept-driven image segmentation. This analysis spans multiple aspects, including network architecture, objective functions, prompt types, training datasets, and evaluation metrics. In [33], SAM2 and SAM3 are further compared for zero-shot segmentation of 3D medical images, with the evaluation restricted to purely geometric prompts and the concept-based mechanism explicitly disabled. The results demonstrate that SAM3 consistently outperforms SAM2 across most medical segmentation tasks, particularly in scenarios involving sparse user annotation or complex anatomical topology. However, the potential of SAM3 for lesion segmentation remains insufficiently explored. Additionally, to improve the domain adaptation capability, SAM3-Adapter introduces lightweight adapter modules into the vision encoder blocks and fine-tunes both the adapter sets and the mask decoder of SAM3, resulting in improved segmentation accuracy, robustness, and training efficiency compared with prior SAM-based adaptation strategies [34]. Concurrently, MedSAM3 [35] extends SAM3 to medical image and video segmentation by leveraging open-vocabulary medical concepts as textual prompts. This approach fine-tunes the detector components while freezing the image and text encoders of the original SAM3 architecture, using paired medical images and corresponding labels. Experimental results demonstrate the effectiveness of this strategy in adapting SAM3 to medical imaging tasks.

Although a series of studies have been conducted on SAM3, including performance comparisons with SAM2 and fine-tuning using adapters, several important questions remain insufficiently explored in the context of concept-prompt-based (text and/or image exemplars) medical image segmentation: (1) how well SAM3 generalizes to tumor or lesion[1] segmentation across different anatomical regions, given the substantial variations in anatomical structures, imaging modalities, data distributions, and shape characteristics (as illustrated in **Fig. 2**); (2) whether integrating prior/additional knowledge (e.g., prediction results from adjacent slices or multi-modality imaging concatenation) into the input can effectively improve segmentation performance; and (3) how to fine-tune SAM3 in an effective and efficient manner for tumor/lesion segmentation in medical images. This study seeks to address these three questions by focusing on *lesion segmentation across multiple anatomical regions*, *prior-knowledge-enhanced input representations*, and *fine-tuning strategies for SAM3 under concept prompts*.

The main contributions of this study are summarized as follows:

(1) We systematically evaluate SAM3 for tumor/lesion segmentation across multiple anatomical regions (brain, lung, liver, pancreas, kidney, skin, etc.) using concept-level prompts on multiparametric MRI, CT, ultrasound, dermoscopy, and endoscopy images (see **Fig. 2** for illustration).

(2) We investigate the impact of integrating prior/additional knowledge into SAM3, including the use of adjacent prediction masks, multi-modality image fusion, and prior annotations.

(3) We conduct a comprehensive comparison of different fine-tuning strategies for SAM3, including partial module fine-tuning, adapter-based fine-tuning, and full model fine-tuning, under various optimization objectives.

---

[1] In this study, we adopt the term "lesion" to denote a broad category of abnormal tissue regions, including tumors, polyps, ischemic stroke lesions, and other pathological findings across different organs, tissue regions, and conceptual regions like lymph node zones.

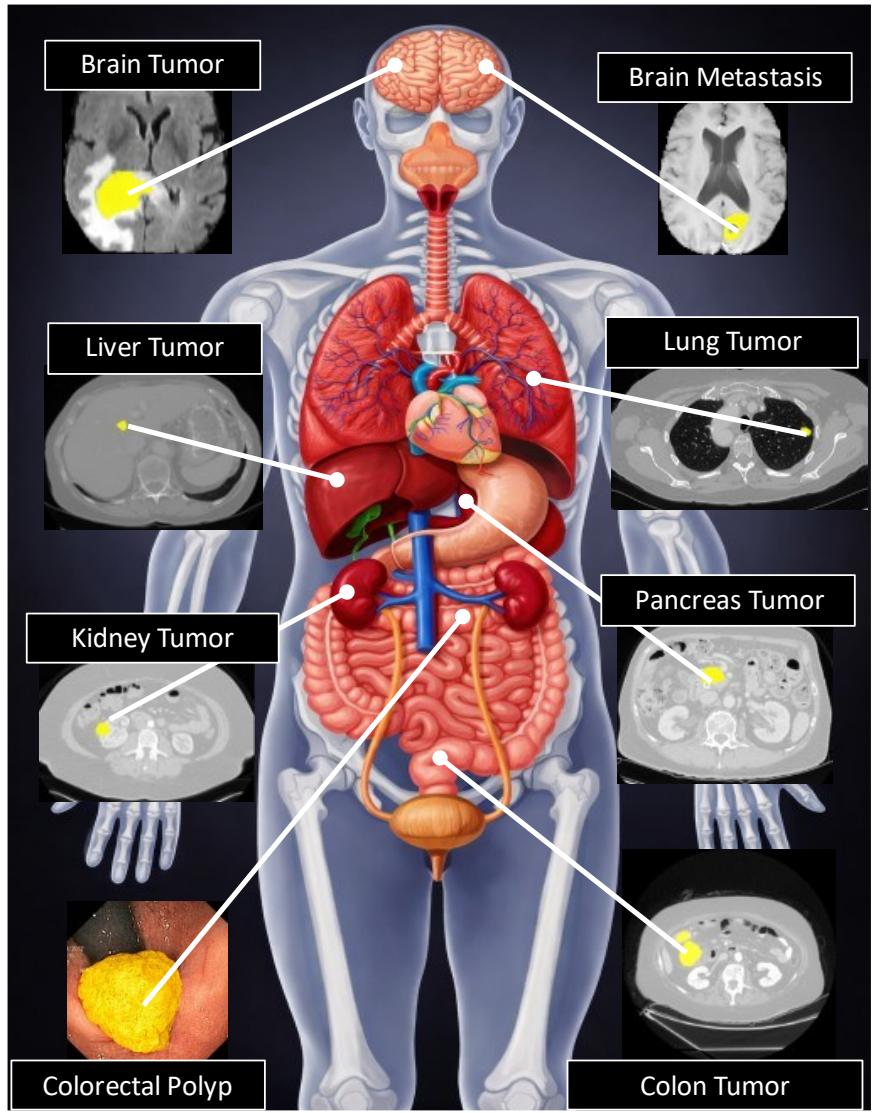

**Figure 2.** Representative examples of tumor and lesion segmentation overlays across multiple medical imaging modalities across diverse anatomical regions. In this study, we focus on the systematic evaluation of a single model's capability to segment heterogeneous lesion types across varied clinical scenarios. The human anatomy illustration is based on a free graphic obtained from Vecteezy.com and was further modified using ChatGPT (https://chatgpt.com/) to enhance visualization.

## 2. Related work

(1). Lesion Segmentation in Medical Images

Lesion segmentation plays a pivotal role in clinical diagnosis, treatment planning, and disease monitoring. Compared with anatomical structure segmentation, lesion segmentation is considerably more challenging due to the large variability in lesion size, shape, appearance, and location, as well as the often low contrast between lesions and surrounding tissues or organs [36, 37]. These factors significantly complicate robust and generalizable model design across different clinical scenarios.

Over the past decade, substantial progress has been achieved in lesion segmentation, driven primarily by two key factors. *First*, a growing number of large-scale, publicly available datasets have been curated and released, largely through community efforts and grand challenges. Representative examples include the Multimodal Brain Tumor Segmentation (BraTS) benchmark, which has evolved from early editions with a limited number of cases to BraTS 2023 [38], comprising 1,424 expertly annotated subjects. Similar efforts include the Kidney Tumor Segmentation (KiTS) dataset for kidney tumors [39], the liver tumor segmentation (LiTS) dataset for liver tumors [40], and the Medical Segmentation Decathlon (MSD), which provides ten diverse organ and tumor segmentation tasks. In addition, modality-specific datasets such as BUSI [41], ISIC2018 [42], and Kvasir-SEG [43] have further advanced research in breast tumor, skin lesion, and polyp segmentation, respectively. *Second*, significant methodological advances have been enabled by deep learning-based segmentation frameworks. Convolutional neural network architectures, particularly encoder-decoder designs such as the U-Net [3] and its variants [36, 44], have become the dominant paradigm due to their strong representation capacity and efficiency. More recent approaches have incorporated attention mechanisms, Transformer-based modules, and hybrid CNN-Transformer architectures to better capture long-range dependencies and contextual information, such as TransUNet [10], UNETR++ [13], and Mask2Former [45, 46]. Despite their success, these methods are typically trained for specific lesion types, anatomical regions, and imaging modalities, limiting their generalization ability in heterogeneous clinical environments and often requiring task-specific retraining or extensive fine-tuning.

Motivated by these limitations, recent research has shifted toward foundation models and prompt-driven segmentation frameworks that aim to provide more generalizable and flexible solutions, such as MedSAM [26], LeSAM [47] and Medical SAM2 [48]. Nevertheless, existing SAM-based approaches rely heavily on geometric prompts for each target object and exhibit limited generalization capability for lesion segmentation across multiple imaging modalities and whole-body anatomical regions. In this study, we propose to leverage SAM3 to enable robust, concept-driven lesion segmentation across diverse organs, imaging modalities, and clinical scenarios. By doing so, our approach aims to bridge the gap between task-specific segmentation models and practical, scalable clinical deployment.

(2). Concept-based Segmentation

With the rapid advancement of large language models, increasing attention has been devoted to learning joint representations that align language and visual modalities. A seminal contribution in this direction is Contrastive Language-Image Pre-training (CLIP), which learns semantically interpretable visual representations (visual grounding) through large-scale text supervision, enabling natural language to be mapped to visual concepts [49]. Building upon this paradigm, Grounding DINO [50] unifies open-set object detection with grounded pre-training and supports free-form textual prompts as input. Further extending this framework, DINO-X [51] integrates multiple perception heads to jointly support object detection, instance segmentation, and object captioning within a unified architecture.

SAM3 adopts pre-trained text and vision encoders from the Perception Encoder (PE) [52] and combines them with a DETR-style detection head and mask decoder for instance-level segmentation. In addition, similar to SAM2, SAM3 incorporates a tracker with a memory bank to enable object tracking and temporal mask propagation. Leveraging its text encoder and prompt encoder, SAM3 supports concept-driven segmentation, allowing all object instances corresponding to a given textual description or image exemplar to be segmented simultaneously. In this work, we extend this capability to the more challenging problem of lesion segmentation across multi-modality medical images, encompassing diverse anatomical regions at the whole-body level.

(3). Fine-tune strategies for foundation models

Training a foundation medical model from scratch is not only computationally expensive, which requires substantial computational resources, but also highly challenging due to the time-consuming nature of collecting and annotating large-scale medical images across multiple modalities, demanding significant domain knowledge. To address these challenges, parameter-efficient fine-tuning (PEFT) methods have been proposed [53].

In general, PEFT approaches can be broadly categorized into two types based on whether additional parameters are introduced into the foundation model: parameter-selection methods and parameter-addition methods. Representative examples of the former include FishMask [54] and BitFit [55], while the latter encompasses adapter-based methods [56] and low-rank adaptation (LoRA) techniques [57]. Building upon the parameter-addition strategies, a series of adapter-based approaches have been developed for fine-tuning SAM-based models, such as SAM2-Adapter [58], 3DSAM-Adapter [59], and DD-SAM2 [29]. In this study, we evaluate different fine-tuning strategies for SAM3 on their effectiveness in medical image lesion segmentation tasks.

## 3. Methodology

### 3.1. Model Overview

SAM3 is designed to support not only geometric prompt-based segmentation (e.g., points, bounding boxes, or masks), as in SAM and SAM2, but also concept prompt-based segmentation using noun phrases or image exemplars. It adopts a dual encoder-decoder Transformer architecture, in which two encoders independently process visual and textual inputs, while two specialized Transformer-based decoders are employed for object localization and pixel-wise segmentation, respectively (see **Fig. 3** for an illustration).

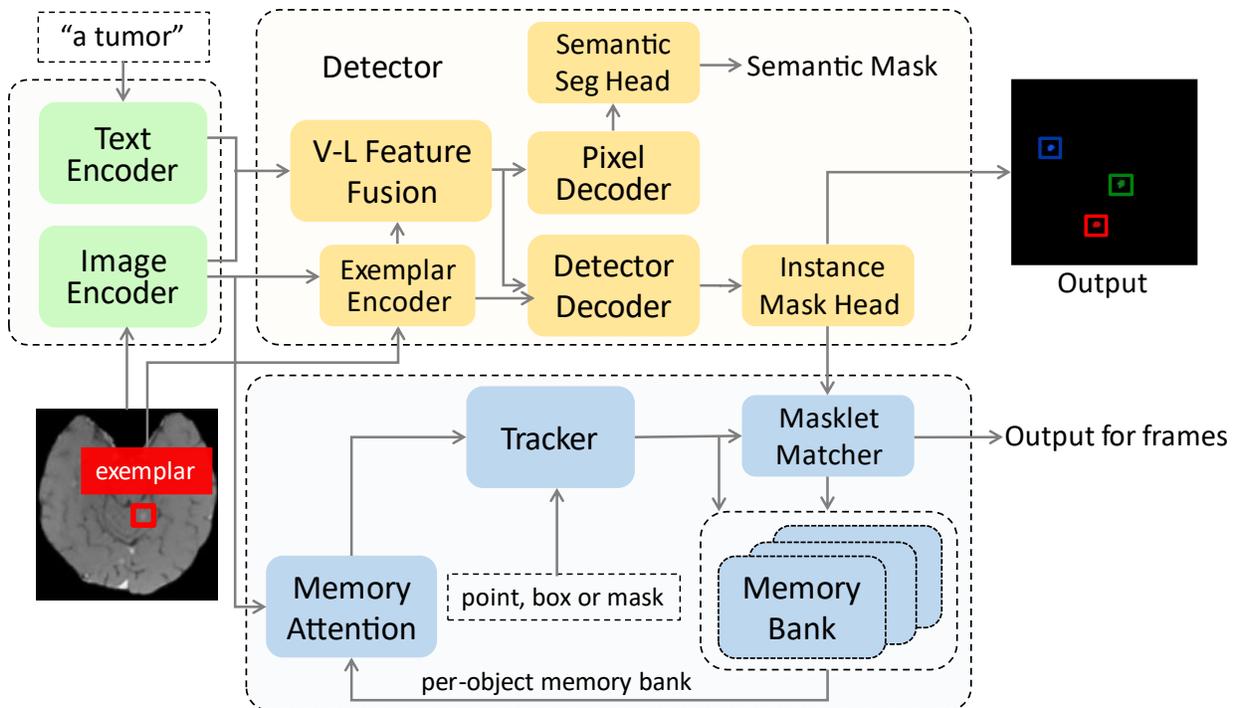

**Figure 3.** Overview of the SAM3 architecture. The model primarily adopts a dual encoder-decoder Transformer framework coupled with a tracker. Specifically, the Text Encoder and Image Encoder comprise approximately 450M and 300M parameters, respectively, while the Detector and Tracker together account for about 100M parameters. In this study, we focus on concept-based, image-level tumor segmentation; therefore, all object tracking components are disabled, as indicated in blue in the figure. For clarity, several skip connections, such as those linking the visual features from the image encoder to the pixel decoder and the text features from the text encoder to the detector decoder, are omitted to emphasize the core structure of SAM3.

Specifically, SAM3 mainly comprises an image encoder, a text encoder, a vision-language feature fusion module[2], a pixel decoder, and an object detection decoder for image-level, concept prompt-based segmentation. In addition, following SAM2, it is coupled with a tracker and a memory bank to support video object segmentation and tracking (In this study, we disable the video tracking functionality and focus exclusively on image-level segmentation). The image encoder and text encoder are inherited from the Perception Encoder backbone [52] to extract image and text embeddings, respectively. The pixel decoder generates dense pixel-level features for semantic mask prediction, whereas the object detection decoder follows a DETR-style design [60], aiming to search for, localize, and identify object instances via learnable detector queries and a presence token. This architecture explicitly decouples detection, segmentation, and tracking objectives, thereby enabling robust concept-driven and multi-instance segmentation.

---

[2] In the original SAM3 supplementary material, the module responsible for fusing visual and textual features is referred to as the Multimodal Decoder. In this work, we adopt the term vision-language fusion module (V-L Feature Fusion) to maintain consistency with the dual encoder-decoder nomenclature used throughout the paper and to better emphasize its architectural role and functional purpose within the dual encoder-decoder framework.

*3.2. Concept-prompt–based segmentation*

In this study, we focus on concept-based, image-level lesion segmentation, following a novel Promptable Concept Segmentation (PCS) paradigm, in which either text or image exemplars are provided as prompts. This contrasts with the Promptable Visual Segmentation (PVS) approach (geometric prompts) employed by SAM and SAM2, although SAM3 also supports PVS segmentation. For text-based prompts, we enable the use of text descriptions to specify the instances to be segmented according to a given concept ("lesions"). The model employs a pre-trained text encoder to extract text embeddings, which are then fed into the vision-language feature fusion module and the detector decoder, facilitating instance recognition and localization.

In addition, SAM3 supports image exemplars as concept prompts. For example, given a bounding box and a positive binary label (exemplar in **Fig. 3**), the model is informed about the target instance to segment. The exemplar encoder processes the geometric information, label class (positive or negative), and the region-of-interest (RoI)-pooled visual features extracted by the image encoder. The detector then integrates these exemplar features to match and identify all instances corresponding to the concept (e.g., the three tumors in **Fig. 2**), in contrast to geometric prompts used in PVS, where each prompt corresponds to only a single segmented object. Furthermore, if text and image exemplars are provided simultaneously, SAM3 can concatenate the text embeddings with the exemplar feature embeddings to generate unified prompt tokens.

*3.3. Prior knowledge enhancement for input*

(1) Enhancement from Preceding Prediction Masks

For volumetric medical images, lesions typically extend across multiple adjacent slices. Exploiting this spatial continuity as prior knowledge, we propose a simple yet effective strategy to integrate information from preceding predictions into the current inference step. Specifically, the prediction mask from the preceding slice is concatenated with the current slice and provided as an additional input channel to the model. This strategy supplies coarse localization cues for the current slice. Importantly, the design of presence tokens and learnable detector queries in the object detection decoder enables SAM3 to distinguish whether a lesion exists at the location suggested by the preceding prediction and to accurately match the corresponding region in the current slice when present. As a result, the model can robustly leverage prior contextual information while avoiding error propagation across slices.

The overall inference procedure can be summarized in the following pseudo code:

```
def prior_enhanced_inference(img_vol, prompt, SAM3):
    """
    Prior-enhanced slice-wise inference for volumetric lesion segmentation.
    Args:
        img_vol:  Tensor of shape [1, N, H, W], input image volume
        SAM3:     Trained SAM3 model
    Returns:
        pred_vol: Tensor of shape [1, N, H, W], predicted segmentation volume
    """
```

```python
    # Initialize prediction volume
    pred_vol = torch.zeros_like(img_vol)

    for n in range(img_vol.shape[1]):
        # Extract current slice: [1, 1, H, W]
        current_slice = img_vol[:, n, :, :].unsqueeze(1)

        if n == 0:
            # No prior mask available for the first slice
            preceding_mask = torch.zeros_like(current_slice)
            model_input = torch.cat(
                (current_slice, current_slice, current_slice), dim=1
            )
        else:
            # Concatenate preceding prediction as prior knowledge
            model_input = torch.cat(
                (current_slice, current_slice, preceding_mask), dim=1
            )

        # Run SAM3 inference
        pred_mask = SAM3(model_input, prompt)

        # Store prediction and update prior
        pred_vol[:, n, :, :] = pred_mask.squeeze(1)
        preceding_mask = pred_mask

    return pred_vol
```

(2) Multimodal Integration as Model Inputs

Multi-modality imaging offers complementary anatomical and functional information that can facilitate more reliable lesion delineation. We therefore explore a multi-modality integration strategy in which images from different modalities are concatenated as multi-channel inputs to SAM3. This input-level fusion enables the model to exploit cross-modal contrast and structural correlations, potentially improving robustness to appearance variability and ambiguous boundaries.

(3) Enhancement from Previous Images and Annotations

In treatment regimens like adaptive radiation therapy (ART) [61], prior treatment images and their corresponding annotations provide valuable contextual information regarding lesion location and morphology. To exploit this prior knowledge, we incorporate historical annotations and the corresponding MR by concatenating them with the current MR image along the channel dimension. Specifically, for each treatment fraction, the MR image and tumor annotation from the immediately preceding treatment fraction are used as auxiliary inputs to guide the segmentation of the current fraction. Leveraging the most recent prior enables the model to capture relevant anatomical and morphological characteristics, which is particularly important when tumor geometry evolves across fractions. This strategy offers coarse spatial guidance that facilitates lesion localization and improves segmentation accuracy. Similar to Sec. 2.3.1, the previous MR image and its corresponding segmentation mask are concatenated with the current MR image as input, allowing the model to directly exploit prior spatial information for tumor detection and delineation. The three prior-based enhancements in Sec. 2.3 are evaluated separately without joint evaluation, to better characterize each strategy's impact on the segmentation accuracy.

*3.4. Strategies for Fine-Tuning SAM3*

For concept-based image-level segmentation, SAM3 involves three core components: the image encoder, the text encoder, and the detector module, as illustrated in **Fig. 2** (highlighted in green and yellow). In this study, we investigate three fine-tuning strategies to adapt SAM3 for lesion segmentation in medical images, namely full fine-tuning, partial fine-tuning, and adapter-based fine-tuning, as summarized in **Table 1**. Note that all tracking-related modules are disabled, as our focus is on image-level segmentation rather than video object tracking and segmentation.

**Table 1.** Three fine-tuning strategies for SAM3.

| Mode | Fine-tuning Module | Description |
|---|---|---|
| Full fine-tuning | Image encoder, text encoder, and detector | End-to-end fine-tuning of all SAM3 components (excluding the tracker). |
| Partial fine-tuning | Detector | Fine-tuning the entire detector module while freezing the image and text encoders. |
| Adapter-based fine-tuning | Adapters in the image encoder | Parameter-efficient fine-tuning via lightweight adapter modules inserted into the image encoder, while freezing the original image encoder, the text encoder, and the detector. |

For the full fine-tuning mode (see **Fig. 4 (a)**), all parameters of the image encoder, the text encoder, and the detector are updated during training. To supervise concept-level pixel predictions without relying on explicit instance matching, two pixel-wise semantic segmentation losses (focal loss and Dice loss) are employed. This design enables effective promptable concept segmentation under pixel-level mask supervision. In addition, a detection objective incorporates multiple instance-level loss terms, including cross-entropy loss for query-level classification, alignment loss, and instance-level mask loss, which enforces accurate spatial alignment between predicted masks and matched ground-truth instances. Together, this dual-level optimization strategy allows SAM3 to jointly model concept presence, spatial extent, and instance specificity, thereby supporting robust and precise lesion segmentation.

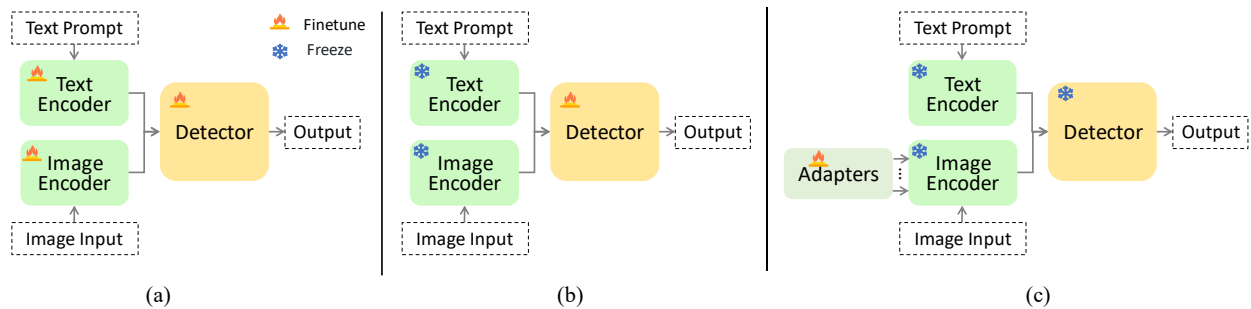

**Figure 4.** (a) Full, (b) partial, and (c) adapter-based fine-tuning strategies.

For partial module fine-tuning (**Fig. 4 (b)**), both the text encoder and image encoder are frozen, and only the detector is fine-tuned. This configuration aims to preserve the general representation capability learned from large-scale pre-training while adapting the task-specific detection head to lesion segmentation. By limiting the number of trainable parameters, this strategy also reduces the risk of overfitting and lowers computational and memory costs. However, freezing the encoders constrains the model's ability to adapt to domain-specific appearance variations in medical images, which may limit performance gains. Therefore, we adopt an adapter-based fine-tuning approach (see **Fig. 4 (c)**) as another configuration for adapting SAM3. Following the pipeline of DD-SAM2 [29], adapter modules are inserted into the blocks of the image encoder. During training, only the parameters of adapter modules are updated, enabling parameter-efficient adaptation while preserving the pretrained backbone weights. For finetuned SAM3, when fine-tuning either all parameters or only the detector module, the learning rate was set to 1e-5, the model was trained for 10 epochs, and the AdamW optimizer was used. The final model was selected based on the highest Dice score on the validation set. For adapter-based fine-tuning, the learning rate was increased to 5e-5, and 12 adapter modules were inserted into the $20^{th}$ to $31^{st}$ blocks of the image encoder. All other training settings remained identical to those described above.

## 4. Experimental Results

*4.1 Datasets*

To comprehensively evaluate the performance of SAM3 across diverse medical imaging modalities, we curated 13 lesion segmentation datasets spanning MRI, CT, dermoscopy, endoscopy, and ultrasound images. These datasets cover a wide range of anatomical regions, including the brain, thorax, abdomen, and skin, and encompass various pathological targets, including brain tumors, lung tumors, liver tumors, pancreatic tumors, kidney tumors, ischemic stroke lesions, skin lesions, and others. A detailed summary of the datasets, imaging modalities, target objects, and data splits (for BraTS2023-MET only) is provided in **Table 2**. For evaluations using a single modality as input, T1-contrast (T1-C) images from BraTS2023-MET and DWI from ISLSE22 are used as the input. For the Mix-Seq-Brain and Mix-Seq-Abdomen datasets, the input imaging modality is randomly selected for each case to evaluate the model's generalization capability across multi-parametric MRI sequences.

**Table 2.** Summary of the medical datasets used for lesion segmentation.

| Dataset | Modality | Target Objects | Train | Val | Test |
| --- | --- | --- | --- | --- | --- |

| Dataset | Modality | Target | Train | Val | Test |
|---|---|---|---|---|---|
| BraTS2023-MET | T1-C, T1-W, T2-FLAIR, T2-W | Brain metastasis | 85 | 80 | 73 |
| ISLSE22 | DWI, ADC | Ischemic stroke lesion | - | - | 125 |
| MSD Lung | CT | Lung tumor | - | - | 30 |
| MSD Liver | CT | Liver tumor | - | - | 60 |
| MSD Pancreas | CT | Pancreas tumor | - | - | 180 |
| MSD Colon | CT | Colon tumor | - | - | 66 |
| KiTS2019 | CT | Kidney tumor | - | - | 200 |
| ISIC2018 | Dermoscopy | Skin lesion | - | - | 1520 |
| Kvasir-SEG | Endoscopy | Polyp | - | - | 320 |
| BUSI | Ultrasound | Breast tumor | - | - | 300 |
| Mix-Seq-Brain | T1-C, T1-W, T2-FLAIR, T2-W | Brain tumor | - | - | 86 |
| One-Seq-Liver | T2-MVXD | Liver tumor | - | - | 115 |
| Mix-Seq-Abdomen | T2-MVXD, T2-MVXD-SPIR, T1-3DVaneXD, and T2-FS | Abdomen tumor | - | - | 88 |

(1) **BraTS2023-MET** [62]. Brain metastases represent the most common malignant tumors of the adult central nervous system and arise from primary cancers located elsewhere in the body. BraTS2023-MET is a subtask of the BraTS2023 challenge, designed to segment brain metastases using multi-modal MRI, including T1-contrast-enhanced (T1-C), T1-weighted (T1-W), T2-FLAIR (T2-F), and T2-weighted (T2-W) images. In this study, we focus on enhancing tumor segmentation and partition the 238 annotated cases into 85 training, 80 validation, and 73 testing subjects.

(2) **ISLSE22** [63]. ISLES22 is a multi-center brain MRI dataset developed to benchmark segmentation algorithms for acute to subacute ischemic stroke lesions. The dataset exhibits substantial variability in lesion size and appearance across diffusion-weighted imaging (DWI), apparent diffusion coefficient (ADC), and FLAIR sequences. We use 125 3D cases, including DWI and ADC volumes with corresponding reference lesion labels, for testing.

(3) **MSD Lung** [64]. The MSD Lung dataset is a subset of the Medical Segmentation Decathlon (MSD) and consists of 3D CT volumes acquired from patients with non-small cell lung cancer. The task focuses on the segmentation of relatively small lung tumors. In our experiments, we use 30 testing samples.

(4) **MSD Liver** [64]. The MSD Liver dataset contains contrast-enhanced CT scans with annotations for liver tumors. In this work, we employ 60 publicly available annotated cases for testing.

(5) **MSD Pancreas** [64]. The MSD Pancreas dataset comprises CT scans from patients undergoing surgical resection for pancreatic masses. The dataset poses a challenging segmentation problem due to severe class imbalance among background, pancreas, and tumor regions. We utilize 180 testing samples, focusing exclusively on pancreatic tumor segmentation.

(6) **MSD Colon** [64]. The MSD Colon dataset is another subset of the Medical Segmentation Decathlon and focuses on colon cancer segmentation from contrast-enhanced CT images. The dataset includes volumetric scans with pixel-wise annotations of colon tumors, which exhibit considerable variability in size, shape, and location. We employ 66 CT volumes in our experiments for testing.

(7) **KiTS2019** [39]. The Kidney Tumor Segmentation 2019 (KiTS2019) dataset consists of contrast-enhanced CT scans, with expert annotations of renal tumors. 200 annotated scans are used in this study as testing cases.

(8) **ISIC2018** [42]. The ISIC2018 dataset is part of the International Skin Imaging Collaboration (ISIC) Challenge for benchmarking skin lesion segmentation algorithms. It consists of dermoscopic images with expert-annotated lesion masks. In this study, 1,520 testing images are used.

(9) **Kvasir-SEG** [43]. Kvasir-SEG contains endoscopic images of colorectal polyps, each accompanied by pixel-wise ground-truth annotations. In this study, 320 testing samples are used.

(10) **BUSI** [41]. The Breast Ultrasound Images Dataset (BUSI) comprises ultrasound images categorized into normal (133), benign (437), and malignant (210) cases. As normal images do not contain lesion annotations, they are excluded from this study. 300 lesion-bearing images are used as testing samples.

(11) **Mix-Seq-Brain**. The in-house Mix-Seq-Brain dataset consists of 86 MRI scans collected from 37 patients with brain tumors treated by a 1.5 T MR-LINAC system (Unity, Elekta AB, Stockholm, Sweden), with 1-5 adaptive fractions per patient. The dataset includes multiple T1- and T2-weighted MRI sequences, such as T2-FLAIR, T1-weighted, and contrast-enhanced T1-weighted imaging. In total, 97 clinical target volumes (CTVs) were manually delineated by expert radiation oncologists. More details can be found in [65].

(12) **One-Seq-Liver**. The in-house One-Seq-Liver dataset consists of 115 MRI scans acquired from 31 patients diagnosed with liver tumors, all obtained using a consistent T2-weighted MultiVane eXtended Dynamic (T2-MVXD) sequence. All patients were treated on a 1.5 T MR-LINAC system, with 1-7 adaptive fractions per patient [53]. In total, 167 internal target volumes (ITVs) were manually delineated by experienced radiation oncologists during routine clinical practice.

(13) **Mix-Seq-Abdomen**. The in-house Mix-Seq-Abdomen dataset comprises 88 fractional MRI scans from 28 MR-LINAC patients with abdominal malignancies, including liver, pancreatic, and renal tumors. Multiple MRI sequences were employed across different treatment fractions, including T2-MVXD, T2-MVXD with fat suppression (T2-MVXD-SPIR), T1-weighted 3D Vane XD (T1-3D-VaneXD), and T2-weighted fat-suppressed turbo spin echo (T2-FS) imaging. Each patient underwent 2-6 adaptive fractions. A total of 156 ITVs were manually annotated by radiation oncologists for this cohort.

*4.2 Direct Evaluation without Fine-Tuning*

(1) Comparison between SAM2 and SAM3 using bounding boxes as geometric prompts

To compare the lesion segmentation performance of SAM2 and SAM3 using bounding box (bbox)-based geometric prompts, we extracted tight bounding boxes directly from the ground-truth masks for each lesion on every image slice. Specifically, each bounding box was defined by the top, bottom, left, and right boundaries of the corresponding ground-truth lesion mask. Although this strategy is less practical in real clinical workflows by requiring precise, slice-by-slice bounding box annotations, it sets a fair reference for model performance comparison. **Table 3** summarizes the segmentation results in terms of DICE, 95th-percentile Hausdorff distance (HD95), and average surface distance (ASD) across all 13 evaluated datasets.

These results reflect the optimal performance achievable by SAM2 and SAM3 when guided by tight bounding box prompts.

As shown in **Table 3**, SAM3 outperforms SAM2 on 10 of the 13 lesion-related datasets in terms of DICE. Notably, SAM3 achieves substantial performance gains on ISLES22 (+4.41%, in absolute terms of DICE, same in the following), MSD Pancreas (+3.04%), and ISIC2018 (+5.39%). In contrast, SAM2 exhibits marginal advantages on BraTS2023-MET, MSD Lung, and BUSI, with Dice improvements <1%. Overall, these results indicate that SAM3 demonstrates superior lesion segmentation capability compared to SAM2 when evaluated under conventional geometric prompting conditions.

**Table 3.** Comparison of segmentation performance between SAM2 and SAM3, using bounding box prompts for every lesion across multiple medical datasets. (↑ indicates higher is better; ↓ indicates lower is better. The same notation is used in the following tables.)

| Dataset | Method | DICE (%)↑ | HD95 (voxel)↓ | ASD (voxel)↓ |
|---|---|---|---|---|
| BraTS2023-MET | SAM2 | 88.52 | 2.18 | 0.65 |
| | SAM3 | 87.92 | 1.45 | 0.54 |
| ISLSE22 | SAM2 | 81.26 | 1.54 | 0.35 |
| | SAM3 | 85.67 | 1.15 | 0.31 |
| MSD Lung | SAM2 | 87.01 | 2.35 | 0.63 |
| | SAM3 | 86.90 | 2.38 | 0.64 |
| MSD Liver | SAM2 | 84.37 | 4.21 | 1.33 |
| | SAM3 | 86.04 | 3.24 | 1.03 |
| MSD Pancreas | SAM2 | 83.88 | 10.12 | 2.39 |
| | SAM3 | 86.92 | 6.13 | 1.46 |
| MSD Colon | SAM2 | 82.52 | 2.98 | 0.69 |
| | SAM3 | 83.89 | 2.96 | 0.60 |
| KiTS2019 | SAM2 | 90.23 | 4.00 | 1.10 |
| | SAM3 | 91.94 | 3.02 | 0.77 |
| ISIC2018 | SAM2 | 85.43 | 246.11 | 95.00 |
| | SAM3 | 90.82 | 119.78 | 46.25 |
| Kvasir-SEG | SAM2 | 94.33 | 27.78 | 8.66 |
| | SAM3 | 94.52 | 21.82 | 7.22 |
| BUSI | SAM2 | 87.49 | 25.94 | 9.16 |
| | SAM3 | 87.41 | 26.79 | 9.92 |
| Mix-Seq-Brain | SAM2 | 76.15 | 17.05 | 4.65 |
| | SAM3 | 77.56 | 8.45 | 2.74 |
| One-Seq-Liver | SAM2 | 82.19 | 4.11 | 1.17 |
| | SAM3 | 83.28 | 3.90 | 1.06 |
| Mix-Seq-Abdomen | SAM2 | 83.96 | 3.51 | 1.02 |
| | SAM3 | 85.62 | 3.07 | 0.87 |

(2) Performance comparison between concept-based prompts for SAM3

In this setting, we evaluate the performance of SAM3 under concept-based prompting. Specifically, we compare three prompting strategies: pure text (T), single image exemplar (E), and a combination of text and image exemplar (T+E). For simplicity and consistency across datasets, the word "lesion" is used as the text prompt in all experiments. For the image exemplar, a tight bounding box of a lesion is randomly selected from a slice and provided to SAM3 as a visual prompt, guiding the model to identify all similar lesions present on the slice. This image exemplar serves as a reference example of the target concept rather than specifying a particular lesion instance. **Table 4** summarizes the segmentation performance under the three concept-based prompting strategies in terms of DICE, HD95, and ASD.

**Table 4.** Segmentation performance of SAM3 using concept-based prompts for volume medical datasets: text (T), image exemplar (E), and combined text and image exemplar (T+E).

| Dataset | Prompt | DICE (%)↑ | HD95 (voxel)↓ | ASD (voxel)↓ |
|---|---|---|---|---|
| BraTS2023-MET | T | 0.69 | nan | inf |
|  | E | 74.39 | 12.96 | 4.68 |
|  | T + E | 77.29 | 4.24 | 1.06 |
| ISLSE22-DWI | T | 42.46 | nan | inf |
|  | E | 68.44 | 7.29 | 1.96 |
|  | T + E | 64.94 | 6.27 | 1.55 |
| MSD Lung | T | 62.36 | nan | inf |
|  | E | 83.90 | 29.34 | 9.63 |
|  | T + E | 86.97 | 2.28 | 0.76 |
| MSD Liver | T | 23.95 | nan | inf |
|  | E | 75.91 | 24.5 | 9.84 |
|  | T + E | 77.45 | 6.29 | 1.84 |
| MSD Pancreas | T | 3.96 | nan | inf |
|  | E | 82.25 | 23.87 | 7.17 |
|  | T+E | 84.97 | 4.74 | 1.31 |
| MSD Colon | T | 9.17 | nan | inf |
|  | E | 81.87 | 18.88 | 6.39 |
|  | T+E | 83.86 | 2.39 | 0.43 |
| KiTS2019 | T | 8.71 | nan | inf |
|  | E | 79.30 | 82.89 | 37.99 |
|  | T+E | 90.84 | 8.81 | 2.12 |
| Mix-Seq-Brain | T | 11.43 | nan | inf |
|  | E | 73.03 | 13.84 | 3.86 |
|  | T+E | 74.29 | 9.28 | 3.00 |
| One-Seq-Liver | T | 3.92 | nan | inf |
|  | E | 72.93 | 26.48 | 8.52 |
|  | T+E | 76.28 | 5.42 | 1.60 |
| Mix-Seq-Abdomen | T | 8.94 | nan | inf |
|  | E | 79.17 | 9.76 | 2.38 |
|  | T+E | 80.01 | 5.73 | 1.40 |

As shown in **Table 4**, text-only prompts (T) consistently yield the poorest segmentation performance across all evaluated datasets, indicating that relying solely on textual descriptions is insufficient for accurate lesion recognition and delineation. For instance, the DICE scores obtained with text-only prompts are extremely low on several datasets, including 0.69% on BraTS2023-MET, 3.96% on MSD Pancreas, 9.17% on MSD Colon, and 1.99% on KiTS2019, highlighting the inherent difficulty of localizing lesions without visual guidance. In contrast, image exemplar-based prompting (E) substantially improves segmentation performance across all datasets. Moreover, the combined use of text and image exemplars (T+E) generally achieves the best or near-best performance, outperforming image exemplars alone in most cases. Specifically, the DICE score is further improved by 2.90%, 3.07%, 2.72% and 11.54% on the BraTS2023-MET, MSD Lung, MSD Pancreas, and KiTS2019 datasets, respectively. The only exception is observed for ischemic stroke lesion segmentation on the ISLES22-DWI dataset, where adding text prompts leads to a slight degradation in DICE.

Notably, despite this decrease in overlap accuracy, the T+E prompting strategy consistently improves surface-based metrics (HD95 and ASD) across all datasets, suggesting that textual prompts contribute to more accurate boundary delineation. Overall, these results demonstrate that visual exemplars play a critical role in concept-based lesion segmentation, while textual prompts provide complementary semantic guidance when used in conjunction with image exemplars.

*(3) Prior-driven Input Augmentation*

a) Input augmentation with preceding slices' prediction masks

**Table 5** presents the results of incorporating the prediction masks from adjacent, preceding slices as an additional input channel. This strategy aims to provide coarse localization priors under the assumption that lesions typically span multiple adjacent slices. As shown in **Table 5**, integrating the previous prediction mask leads to DICE improvements on several datasets, for instance BraTS2023-MET, Mix-Seq-Brain, and Mix-Seq-Abdomen, demonstrating that even simple mask-based priors can be beneficial for lesion segmentation.

However, the effectiveness of this strategy is not consistent across all datasets. For example, performance degradation is observed on MSD Lung, and metric trade-offs are evident on Mix-Seq-Brain, where DICE improves substantially while the surface-based metric HD95 deteriorates. These results suggest that naïvely concatenating preceding prediction masks may introduce noisy or misleading spatial cues, particularly when lesions exhibit discontinuous appearance across slices, significant shape variation, or when prediction errors propagate from preceding slices. More robust strategies, such as confidence-aware mask fusion or selective gating mechanisms, may be necessary to fully exploit mask-based priors while avoiding error accumulation and false-positive propagation.

**Table 5.** Quantitative comparison of segmentation performance with and without using preceding slices' prediction masks as prior, for four public and three in-house datasets. Results are reported for models using text and image exemplar prompts.

| Dataset | Preceding Slice's Prediction Mask | DICE (%)↑ | HD95 (voxel)↓ | ASD (voxel)↓ |
|---|---|---|---|---|
| BraTS2023-MET | w/o | 77.29 | 4.24 | 1.06 |
|  | w | 78.11 | 3.20 | 0.88 |
| MSD Lung | w/o | 86.97 | 2.28 | 0.76 |
|  | w | 84.15 | 3.23 | 1.02 |
| KiTS2019 | w/o | 90.84 | 8.81 | 2.12 |
|  | w | 90.33 | 4.62 | 0.86 |
| Mix-Seq-Brain | w/o | 74.29 | 9.28 | 3.00 |

|  | w | 78.24 | 9.79 | 2.86 |
| --- | --- | --- | --- | --- |
| One-Seq-Liver | w/o | 76.28 | 5.42 | 1.60 |
|  | w | 76.29 | 10.56 | 1.05 |
| Mix-Seq-Abdomen | w/o | 80.01 | 5.73 | 1.40 |
|  | w | 80.74 | 8.35 | 1.15 |

b) Input augmentation with multimodal images

To assess the effectiveness of multi-modality input augmentation, we evaluate our approach on two multimodal datasets: BraTS2023-MET and ISLSE22. As summarized in **Table 6**, incorporating multiple imaging modalities generally leads to improved segmentation performance, indicating that complementary modality-specific information is beneficial for lesion delineation.

On the BraTS2023-MET dataset, progressively adding imaging modalities consistently improves performance. In particular, combining T1-C, T2-F, and T2-W images yields a 2.78% absolute DICE improvement compared with using only the T1-C modality. These results demonstrate that multimodal fusion effectively enriches lesion-related contextual information and enhances segmentation robustness. For the ISLSE22 dataset, while single-modal performance varies across DWI and ADC inputs, their joint use results in improved surface-based metrics (HD95 and ASD), highlighting the complementary nature of diffusion-derived modalities. However, the DICE of the combined input improves modestly over ADC-only results, while being inferior compared with the DWI-only inputs. It suggests that modality relevance and quality also play a critical role in determining the effectiveness of multimodal fusion. Overall, these findings indicate that input-level multi-modality concatenation is a simple yet potentially effective strategy for improving lesion segmentation performance, although its benefits may be dataset- and modality-dependent.

**Table 6.** Quantitative evaluation of multimodal input augmentation on BraTS2023-MET and ISLSE22 datasets. Results are reported for models using text and image exemplar prompts.

| Dataset | Input | DICE (%)↑ | HD95 (voxel)↓ | ASD (voxel)↓ |
| --- | --- | --- | --- | --- |
| BraTS2023-MET | T1-C | 77.29 | 4.24 | 1.06 |
|  | T1-C+T2-F | 77.82 | 3.75 | 0.99 |
|  | T1-C+T1-W | 77.53 | 3.55 | 0.98 |
|  | T1-C+T2-W | 78.61 | 3.22 | 0.89 |
|  | T1-C+T2-F+T1-W | 79.25 | 3.80 | 1.08 |
|  | T1-C+T2-F+T2-W | 80.07 | 4.03 | 1.08 |
| ISLSE22 | DWI | 64.94 | 6.27 | 1.55 |
|  | ADC | 58.03 | 6.24 | 1.43 |
|  | DWI + ADC | 60.09 | 5.26 | 0.79 |

c) Input augmentation by using prior images/annotations

Treatment regimens like adaptive radiation therapy (ART) involve repeated imaging and segmentation for the same patient across multiple sequential treatment fractions, where prior knowledge is usually available to guide each new, patient-specific segmentation task. To leverage this information, we used the three in-house datasets that included both the current MR images and the prior MR images (along with their corresponding tumor annotations) from previous treatment sessions. As described in Sec. 2.3.3, the current MR scan is supplemented with prior MR images and their annotations, resulting in a three-channel input that encodes both temporal and anatomical context for segmentation. Quantitative results are summarized in **Table 7**. Incorporating prior MR information yields consistent performance gains, achieving absolute Dice improvements of 10.19%, 7.87%, and 4.47% on the Mix-Seq-Brain, One-Seq-Liver, and Mix-Seq-Abdomen datasets, respectively. These findings indicate that the integration of previous MR images and annotations effectively enhances SAM3's performance for lesion segmentation in the ART setting.

**Table 7.** Segmentation performance of SAM3 with and without prior MR images and annotations as guidance. Results are reported for models using text and image exemplar prompts.

| Dataset | Prior image/annotation-guided | DICE (%)↑ | HD95 (voxel)↓ | ASD (voxel)↓ |
|---|---|---|---|---|
| Mix-Seq-Brain | No | 74.29 | 9.28 | 3.00 |
|  | Yes | 85.08 | 6.89 | 1.97 |
| One-Seq-Liver | No | 76.28 | 5.42 | 1.60 |
|  | Yes | 84.15 | 2.89 | 0.74 |
| Mix-Seq-Abdomen | No | 80.01 | 5.73 | 1.40 |
|  | Yes | 84.48 | 3.03 | 0.82 |

*4.3 Evaluation of Fine-Tuning Strategies*

To evaluate the effectiveness of fine-tuning, we fine-tuned SAM3 using the BraTS2023-MET dataset. Three fine-tuning strategies were investigated: fine-tuning all model parameters, fine-tuning only the detector module, and fine-tuning only the additional adapter modules. An overview of these strategies is illustrated in **Fig. 4**.

**Table 8** presents a quantitative comparison of the different fine-tuning strategies on the BraTS2023-MET dataset. Among all approaches, adapter-based fine-tuning achieves the best overall performance, yielding the highest Dice score (82.96%), as well as the lowest average surface distance (ASD). In contrast, fine-tuning all parameters or only the detector module provides limited or inconsistent improvements compared to the baseline without fine-tuning. These results highlight the effectiveness of parameter-efficient adapter-based fine-tuning for lesion segmentation.

**Table 8.** Comparison between different fine-tuning strategies on the BraTS2023-MET dataset.

| FT Module | DICE (%)↑ | HD95 (voxel)↓ | ASD (voxel)↓ |
|---|---|---|---|
| None | 77.29 | 4.24 | 1.06 |
| All | 79.16 | 6.81 | 1.45 |
| Detector | 76.42 | 6.22 | 1.03 |
| Adapter | 82.96 | 5.42 | 0.90 |

*4.4 Visual Results*

(1) Visual results of SAM2 and SAM3 using bounding box prompts

**Figure 5** presents representative lesion segmentation results obtained with SAM2 and SAM3 across multiple datasets, including brain, breast, polyp, skin, lung, liver, pancreas, and kidney, where tight bounding boxes are used as geometric prompts. Overall, both models demonstrate comparable segmentation performance on most samples, indicating similar generalization ability across different anatomical structures and imaging modalities. However, SAM3 exhibits improved contextual understanding in several challenging cases, as reflected in the breast and pancreas examples shown in **Figure 5**. In these instances, SAM3 produces segmentations that better align with lesion boundaries and surrounding anatomical context, suggesting that the model benefits from enhanced feature representation and stronger global reasoning capability compared with SAM2.

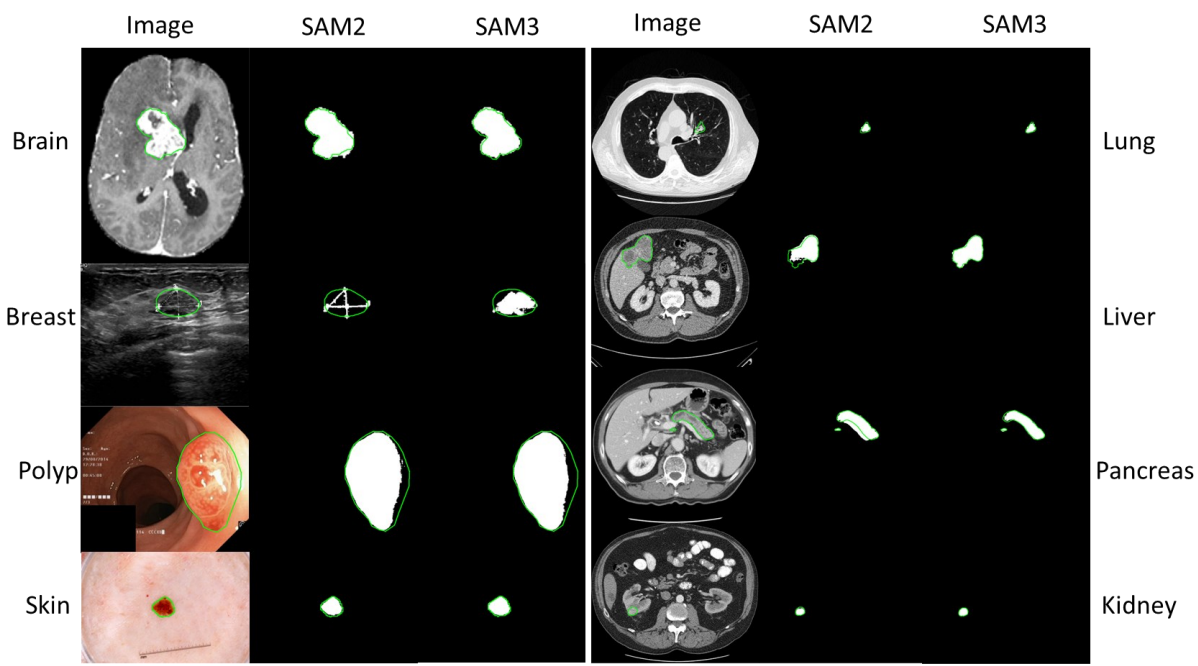

**Figure 5**. Visualization of lesion segmentation results obtained with SAM2 and SAM3 using bounding boxes as prompts. The green contours indicate the ground truth annotations, and the white masks represent the predicted segmentation results. The same color convention is used in the following figures.

(2) Visual results of SAM3 using concept-based prompts

**Figure 6** demonstrates visual results of SAM3 for lesion segmentation using three types of prompts: text description ("lesion"), image exemplar, and the combination of text and image exemplar. Examples from the brain, breast, lung, liver, and pancreas datasets are shown. The results demonstrate that: 1) Using only text prompts yields the poorest performance, as several lesions cannot be correctly detected in multiple cases shown in **Figure 6**; 2) Using image exemplars significantly improves segmentation accuracy; however, it tends to produce over-segmentation in some cases, such as the breast and lung examples; 3) Combining text and image exemplar prompts achieves the best overall performance, producing more accurate lesion localization and boundaries across different organs.

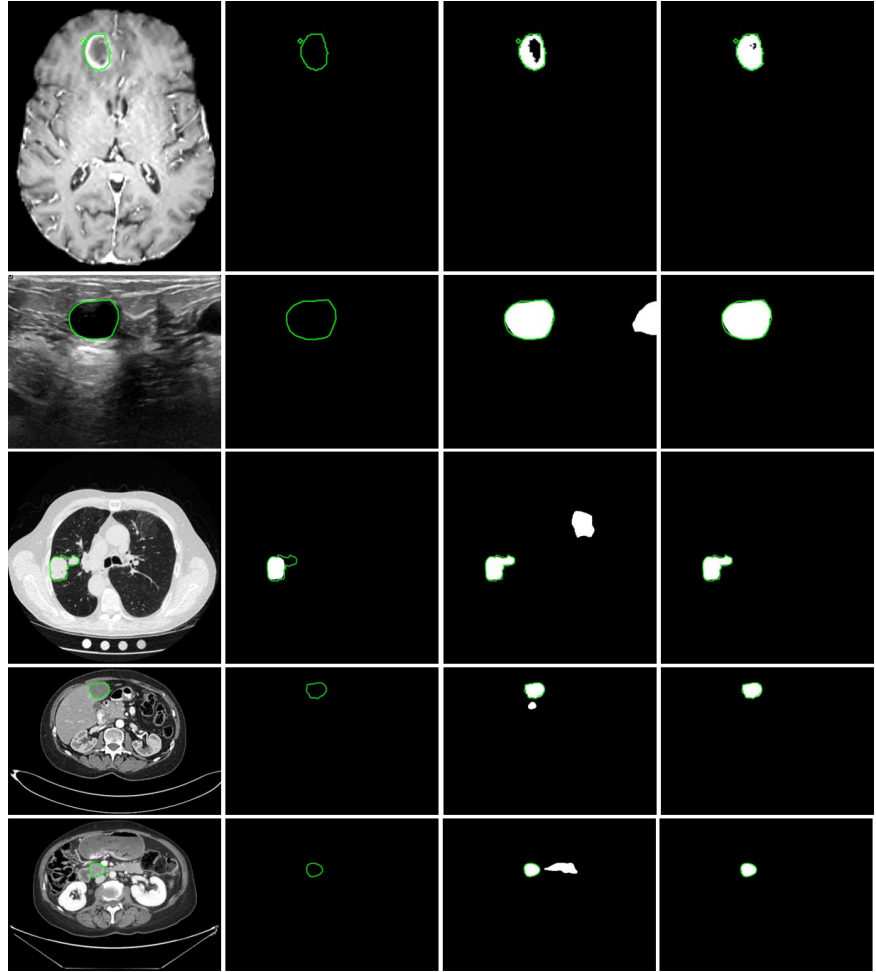

**Figure 6.** Visualization of lesion segmentation results in the brain, breast, lung, liver, and pancreas using different prompt types, including text (2nd column), image exemplar (3rd column), and their combination (4th column).

(3) Visual results of SAM3 with and without using previous images and annotations as prior knowledge

**Figure 7** presents visual results on our in-house adaptive radiation therapy datasets, including Mix-Seq-Brain, One-Seq-Liver, and Mix-Seq-Abdomen, for brain, liver, and abdominal lesion segmentation using a combination of the text and the image exemplar as prompts. As shown in the figure, incorporating previous images and their corresponding annotations as additional guidance consistently improves the final segmentation results. This observation is consistent with the quantitative results reported in **Table 7**, demonstrating that temporal or sequential prior information provides useful contextual cues for more accurate lesion localization and boundary delineation.

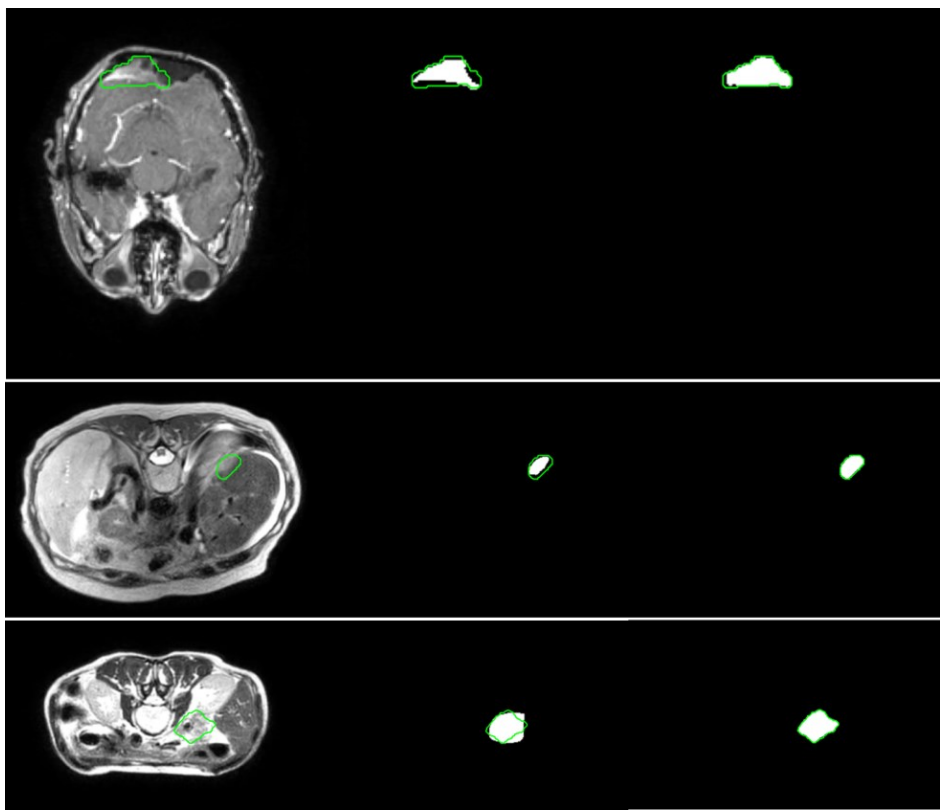

**Figure 7**. Visual segmentation results without prior image and annotation knowledge (2nd column) and with prior knowledge (3rd column).

(4) Visual comparison of segmentation results under different fine-tuning strategies

**Figure 8** presents the visual results obtained with the original SAM3 model, SAM3 with full fine-tuning, SAM3 with detector-only fine-tuning, and SAM3 with adapter-based fine-tuning on the BraTS2023-MET dataset with combined texture and image exemplar as prompts. As shown, fine-tuning consistently improves segmentation performance over the original SAM3 model, indicating the benefit of task-specific adaptation.

Among the three fine-tuning strategies, detector-only fine-tuning yields inferior results compared with both full fine-tuning and adapter-based fine-tuning, suggesting that updating only the detection module is insufficient for optimal lesion segmentation. Furthermore, when comparing full fine-tuning with adapter-based fine-tuning, the adapter-based approach achieves better results on this dataset. This observation suggests that, under limited training data, full fine-tuning may introduce a higher risk of overfitting, while adapter-based fine-tuning provides a more stable and parameter-efficient alternative.

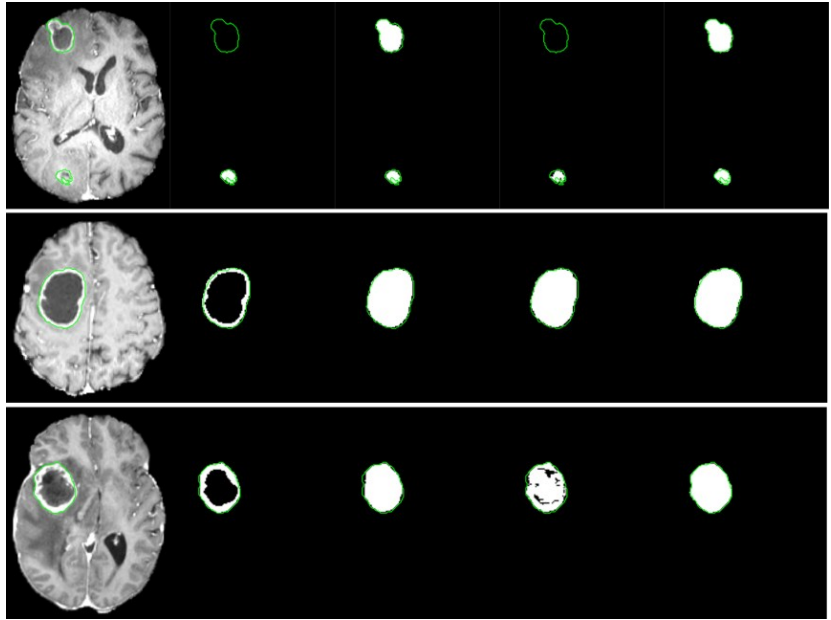

**Figure 8.** Visual segmentation results obtained with the original SAM3 (2$^{nd}$ column), SAM3 with full fine-tuning (3$^{rd}$ column), SAM3 with detector-only fine-tuning (4$^{th}$ column), and SAM3 with adapter-based fine-tuning (5$^{th}$ column). The green contours indicate the ground truth, and the white masks denote the predicted segmentations.

## 5. Discussion

In this study, we evaluated the lesion segmentation performance of the original SAM3 framework using concept-based prompts. We also explored potential strategies to further improve effectiveness through the integration of prior knowledge via input enhancement without additional training. Finally, we investigated multiple fine-tuning strategies for SAM3. Although both prior-knowledge augmentation and fine-tuning improved the segmentation performance of SAM3, several limitations remain and warrant investigations in future work.

*(1) Prior knowledge integration as input*

Prior knowledge, such as spatial location, relative scale, and multimodal imaging information, plays a critical role in accurate lesion localization and delineation. Previous work, including SAM2-Aug [65], has demonstrated that integrating prior MR images and their corresponding annotations into SAM2 can substantially improve tumor localization and segmentation performance in adaptive radiation therapy scenarios, where historical imaging and annotations are readily available. In this study, we observe that a similar strategy is also effective for SAM3 (**Table 7**, **Figure 7**). The effectiveness of this approach is largely attributed to the fact that tumor morphology and location typically exhibit limited variations between consecutive treatment sessions. However, naïvely supplementing prior MR images and their annotations with the current input may be less ideal. Such a strategy primarily provides coarse spatial cues and may be susceptible to additional noise, including inaccurate boundaries and deformation/motion between current and prior imaging.

Moreover, we find that directly using predicted masks from preceding slices of 3D volumes as prior can improve SAM3's segmentation performance in some but not all scenarios (**Table 5**). For example, a previous slice may contain a lesion while the current slice does not, leading to the introduction of misleading information and an increased risk of false-positive predictions. Similarly, using multimodal images as input generally helps with the segmentation performance, but can also lead to decreased accuracy in some scenarios (**Table 6**). These observations highlight a key limitation of naïve prior-knowledge integration.

Therefore, an effective segmentation framework should be capable of distinguishing subtle discrepancies between the current image and auxiliary prior information, while selectively leveraging prior knowledge to refine segmentation results. Learning such adaptive fusion mechanisms remains an important direction for improving the robustness, generalizability, and detection accuracy of SAM3-based lesion segmentation.

*(2) Automatic prompt generation*

Prompt augmentation proves an effective strategy for SAM-based methods [65, 66]. In this study, we investigate its effectiveness for lesion segmentation under concept-based prompting, using text and/or image exemplars. Our initial results indicate that relying solely on text prompts is insufficient for accurate lesion localization, primarily due to the low contrast of lesions and their visual similarity to surrounding tissues or organs. While combining text prompts with image exemplars significantly improves performance, this approach remains inefficient and potentially impractical in clinical settings, as it requires manual identification of lesion exemplars for each image slice.

To address this limitation, a simplified prompt augmentation strategy could be explored, in which one or a small set of image exemplars from a training image or a subset of training images is reused as shared prompts across all test samples in SAM3. Specifically, in a one-shot scenario, a single lesion image exemplar from the training set is selected and reused as a common prompt for all test samples. In a multi-exemplar scenario, a set of lesion exemplars can first be constructed based on predefined selection criteria or similarity metrics. Subsequently, an automatic selection algorithm can be employed to identify one or more representative exemplars to serve as prompts from the library. Compared with per-image or per-slice exemplar prompting, both strategies could substantially improve computational and annotation efficiency while still providing meaningful visual guidance to the model.

In addition, several promising directions merit further investigation for candidate lesion detection using image exemplars as guidance. First, at the image level, anatomical priors could be leveraged to copy and place image exemplars into anatomically corresponding regions using multi-scale and deformable transformations, while also accounting for intensity distribution differences between the exemplar and target images. Second, at the feature level, multi-layer features extracted from the SAM3 image encoder could be exploited to compute similarity between the exemplar and test samples, enabling automatic candidate region or bounding-box generation for adaptive exemplar prompting. Third, it may be beneficial to design and train a dedicated prompt generation module or to leverage the streaming memory mechanism of SAM3, allowing the model to capture temporal relationships across frames and track lesions using a single prompt.

*(3) Fine-tuning for SAM3*

Fine-tuning is essential for mitigating the domain gap between natural images and medical imaging data when adapting foundation models [27, 35]. We investigated different fine-tuning strategies for SAM3 and found the one focused on lightweight adapter modules to be the most effective. While this approach demonstrates performance improvements on one dataset, several limitations remain.

First, the issue of knowledge degradation after fine-tuning warrants further investigation. The performance, especially on generalizability, can drop following fine-tuning, raising concerns about catastrophic forgetting, particularly when fully fine-tuning large pre-trained models. Developing strategies to preserve generalizable knowledge while adapting to domain-specific tasks remains an open challenge. Second, in this study, adapter modules are inserted into the image encoder based on empirical design choices. However, principled guidelines for determining which components of SAM3 should be fine-tuned are still lacking. For example, it remains unclear whether fine-tuning should also involve the prompt encoder, detector, or mask decoder, and how such choices affect performance, stability, and generalization. Finally, jointly exploring fine-tuning and model distillation may offer a promising direction. Such a strategy could not only enhance model generalization and segmentation accuracy but also facilitate the derivation of compact and efficient models suitable for deployment in resource-constrained or edge-computing environments.

*(4) Uncertainty modeling and human-based prior knowledge*

Binary lesion segmentation is often an ill-defined problem. Clinically, lesions can be categorized as conspicuous lesions (CL) and non-conspicuous lesions (nCL) [37]. In this study, we focus on CL, for which binary annotations are typically regarded as the default ground truth. However, in practice, lesion boundaries are rarely crisp or strictly binary. Both CL and nCL often exhibit gradual transitions from regions of high diagnostic confidence to areas of lower certainty. Consequently, modeling the uncertainty or fuzziness of lesion boundaries is meaningful, as it may provide additional information that can assist clinical interpretation and decision-making. Several studies [37, 67, 68] have explored disease quantification without explicit object delineation to improve the efficiency of clinical workflows. In addition, inspired by recent advances in vision-language modeling, clinically relevant knowledge could be further integrated into segmentation frameworks. For example, textual descriptions can capture modality-specific characteristics of lesions and encode expert knowledge regarding their appearance across different imaging modalities. Lesions that appear clearly in CT may present differently in PET, and vice versa. Therefore, incorporating human-based prior knowledge in textual form to describe such cross-modality differences may help improve both segmentation accuracy and efficiency.

## 6. Conclusion

In this study, we present a comprehensive evaluation of the Segment Anything Model 3 (SAM3) for concept-based lesion segmentation across different medical image modalities and diverse anatomical regions. A total of 13 datasets encompassing 11 lesion types were investigated to assess the generalization capability, robustness, and limitations of SAM3 under text- and exemplar-based prompting without reliance on explicit geometric annotations. Our results demonstrate that SAM3 exhibits strong cross-dataset and cross-modality generalization, although performance variability is observed across different lesion types and imaging modalities. By integrating prior knowledge and employing parameter-efficient fine-tuning strategies, segmentation accuracy and robustness can be further improved. These findings suggest that concept-driven vision-language foundation models hold significant promise for scalable lesion segmentation and may serve as a foundation for more flexible and annotation-efficient clinical workflows.

**Acknowledgement**

Authors would also like to sincerely thank Hengrui Zhao, Yan Dai, Yajun Yu, Hao Peng, Steve Jiang, Joe Wise, Charles Gibson, and Lana Shar for their valuable assistance and support.

**Authors' contributions**

Guoping Xu contributed to writing-original draft, methodology, software, visualization, conceptualization, formal analysis, data curation; Jayaram K. Udupa and Yubing Tong contributed to writing-review and editing; Xin Long, Ying Zhang and Jie Deng contributed to conceptualization and data curation; Weiguo Lu contributed to data curation and conceptualization; You Zhang contributed to methodology, supervision, funding acquisition, writing-review and editing.

**Funding**

The study was supported by the US National Institutes of Health (R01 CA240808, R01 CA258987, R01 EB034691, and R01 CA280135).

**Declarations**

**Availability of supporting data**

The datasets used and/or analyzed during the current study are available from the corresponding author on reasonable request.

**Competing interests**

The authors declare that they have no known competing financial interests or personal relationships that could have appeared to influence the work reported in this paper.

42	Codella, N., Rotemberg, V., Tschandl, P., Celebi, M.E., Dusza, S., Gutman, D., Helba, B., Kalloo, A., Liopyris, K., and Marchetti, M.: 'Skin lesion analysis toward melanoma detection 2018: A challenge hosted by the international skin imaging collaboration (isic)', arXiv preprint arXiv:1902.03368, 2019

43	Jha, D., Smedsrud, P.H., Riegler, M.A., Johansen, D., De Lange, T., Halvorsen, P., and Johansen, H.D.: 'Resunet++: An advanced architecture for medical image segmentation', in Editor (Ed.)^(Eds.): 'Book Resunet++: An advanced architecture for medical image segmentation' (IEEE, 2019, edn.), pp. 225-2255

44	Isensee, F., Jaeger, P.F., Kohl, S.A., Petersen, J., and Maier-Hein, K.H.: 'nnU-Net: a self-configuring method for deep learning-based biomedical image segmentation', Nature methods, 2021, 18, (2), pp. 203-211

45	Cheng, B., Misra, I., Schwing, A.G., Kirillov, A., and Girdhar, R.: 'Masked-attention mask transformer for universal image segmentation', in Editor (Ed.)^(Eds.): 'Book Masked-attention mask transformer for universal image segmentation' (2022, edn.), pp. 1290-1299

46	Sheng, J.-C., Liao, Y.-S., and Huang, C.-R.: 'Apply masked-attention mask transformer to instance segmentation in pathology images', in Editor (Ed.)^(Eds.): 'Book Apply masked-attention mask transformer to instance segmentation in pathology images' (IEEE, 2023, edn.), pp. 342-345

47	Gu, Y., Wu, Q., Tang, H., Mai, X., Shu, H., Li, B., and Chen, Y.: 'LeSAM: Adapt Segment Anything Model for medical lesion segmentation', IEEE Journal of Biomedical and Health Informatics, 2024

48	Zhu, J., Qi, Y., and Wu, J.: 'Medical sam 2: Segment medical images as video via segment anything model 2', arXiv preprint arXiv:2408.00874, 2024

49	Radford, A., Kim, J.W., Hallacy, C., Ramesh, A., Goh, G., Agarwal, S., Sastry, G., Askell, A., Mishkin, P., and Clark, J.: 'Learning transferable visual models from natural language supervision', in Editor (Ed.)^(Eds.): 'Book Learning transferable visual models from natural language supervision' (PmLR, 2021, edn.), pp. 8748-8763

50	Liu, S., Zeng, Z., Ren, T., Li, F., Zhang, H., Yang, J., Jiang, Q., Li, C., Yang, J., and Su, H.: 'Grounding dino: Marrying dino with grounded pre-training for open-set object detection', in Editor (Ed.)^(Eds.): 'Book Grounding dino: Marrying dino with grounded pre-training for open-set object detection' (Springer, 2024, edn.), pp. 38-55

51	Ren, T., Chen, Y., Jiang, Q., Zeng, Z., Xiong, Y., Liu, W., Ma, Z., Shen, J., Gao, Y., and Jiang, X.: 'Dino-x: A unified vision model for open-world object detection and understanding', arXiv preprint arXiv:2411.14347, 2024

52	Bolya, D., Huang, P.-Y., Sun, P., Cho, J.H., Madotto, A., Wei, C., Ma, T., Zhi, J., Rajasegaran, J., and Rasheed, H.: 'Perception encoder: The best visual embeddings are not at the output of the network', arXiv preprint arXiv:2504.13181, 2025

53	Zheng, H., Shen, L., Tang, A., Luo, Y., Hu, H., Du, B., Wen, Y., and Tao, D.: 'Learning from models beyond fine-tuning', Nature Machine Intelligence, 2025, 7, (1), pp. 6-17

54	Sung, Y.-L., Nair, V., and Raffel, C.A.: 'Training neural networks with fixed sparse masks', Advances in Neural Information Processing Systems, 2021, 34, pp. 24193-24205

55	Zaken, E.B., Goldberg, Y., and Ravfogel, S.: 'Bitfit: Simple parameter-efficient fine-tuning for transformer-based masked language-models', in Editor (Ed.)^(Eds.): 'Book Bitfit: Simple parameter-efficient fine-tuning for transformer-based masked language-models' (2022, edn.), pp. 1-9

56	Houlsby, N., Giurgiu, A., Jastrzebski, S., Morrone, B., De Laroussilhe, Q., Gesmundo, A., Attariyan, M., and Gelly, S.: 'Parameter-efficient transfer learning for NLP', in Editor (Ed.)^(Eds.): 'Book Parameter-efficient transfer learning for NLP' (PMLR, 2019, edn.), pp. 2790-2799

57	Hu, E.J., Shen, Y., Wallis, P., Allen-Zhu, Z., Li, Y., Wang, S., Wang, L., and Chen, W.: 'Lora: Low-rank adaptation of large language models', ICLR, 2022, 1, (2), pp. 3

58	Chen, T., Lu, A., Zhu, L., Ding, C., Yu, C., Ji, D., Li, Z., Sun, L., Mao, P., and Zang, Y.: 'Sam2-adapter: Evaluating & adapting segment anything 2 in downstream tasks: Camouflage, shadow, medical image segmentation, and more', arXiv preprint arXiv:2408.04579, 2024
30